\documentclass[aps,prl,reprint,amsmath,amssymb,floatfix,citeautoscript,superscriptaddress]{revtex4-1}
\usepackage{amsmath}
\usepackage{graphicx}
\usepackage{amssymb}
\usepackage{amsthm}
\usepackage{bm}

\usepackage{multirow}
\usepackage{times}
\usepackage{txfonts}

\usepackage{footnote}
\usepackage{afterpage}

\usepackage{color}

\let\vr\undefined

\newcommand{\vr}{{\mathbf{r}}}

\newcommand{\vq}{{\mathbf{q}}}

\newcommand{\vrho}{{\bm{\rho}}}

\begin{document}

\title{Theory of Neutral and Charged Excitons in Monolayer Transition Metal Dichalcogenides}

\author{Timothy C. Berkelbach}
\affiliation{Department of Chemistry, Columbia University, 3000 Broadway, New York, New York 10027, USA}

\author{Mark S. Hybertsen}
\affiliation{Center for Functional Nanomaterials, Brookhaven National Laboratory, Upton, New York 11973-5000, USA}

\author{David R. Reichman}
\email{drr2103@columbia.edu}
\affiliation{Department of Chemistry, Columbia University, 3000 Broadway, New York, New York 10027, USA}

\begin{abstract}
We present a microscopic theory of neutral excitons and charged excitons (trions) in 
monolayers of transition metal dichalcogenides, including molybdenum disulfide.  Our theory is
based on an effective mass model of excitons and trions, parametrized by \textit{ab initio} calculations
and incorporating a proper treatment of screening in two dimensions.  The calculated exciton
binding energies are in good agreement with high-level many-body computations based on the 
Bethe-Salpeter equation.  Furthermore, our calculations for the more complex trion species compare very favorably
with recent experimental measurements, and provide atomistic insight into the microscopic features which
determine the trion binding energy.
\end{abstract}

\maketitle

Monolayer transition metal dichalcogenides (TMDs) have recently emerged as two-dimensional (2D) semiconducting alternatives
to metallic graphene with remarkable properties~\cite{nov05,mak10,spl10}.  For example, MoS$_2$, a prototypical 
family member, exhibits strong photoluminescence~\cite{mak10,spl10}, high charge mobility~\cite{rad11}, and selective optical pumping of
spin and valley degrees of freedom~\cite{xia12,zen12,mak12}.  Typically produced by mechanical exfoliation,
MoS$_2$ has more recently been synthesized via chemical vapor deposition~\cite{lee12,van13},
opening the door to further investigation on large, high-quality samples and incorporation into atomically thin
optoelectronic devices. Quite recently,
several studies have demonstrated the generation and electrostatic manipulation of singly-charged
excitons or \textit{trions} in MoS$_2$~\cite{mak13_trion}, MoSe$_2$~\cite{ross13}, and WSe$_2$~\cite{jon13},
akin to those previously observed in quasi-2D semiconductor quantum wells~\cite{khe93,fin95,hua00}.
However, the large magnitude of observed trion binding energies (20--30 meV) is unprecedented.
This is a clear signal that such atomically thin
semiconductors exhibit unusually strong Coulomb interactions.

\begin{figure}[b]
\centering
\includegraphics[scale=1.0]{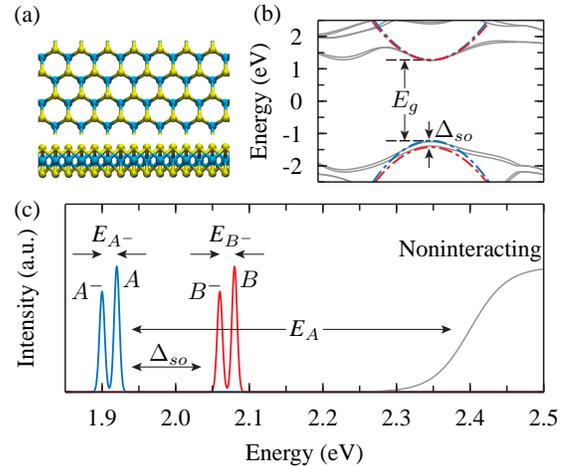}
\caption{
Top and side view of monolayer MoS$_2$ (a), low-energy band structure near the $K$ point calculated by density functional theory
with a rigid shift to increase the gap (b), and schematic absorption spectrum (c). Panel
(b) shows the parabolic band structure assumed in the effective mass approximation for the $A$ (blue) and $B$ (red) excitons that result
from the sizable spin-orbit splitting $\Delta_{so} \approx$ 200 meV.
}
\label{fig:scheme}
\end{figure}

The scenario revealed by experiments to date is summarized pictorially in Fig.~\ref{fig:scheme},
which shows the 2D hexagonal lattice of MoS$_2$ in panel (a) and the low-energy band structure near the fundamental, direct gap at the $K$ point in panel (b),
including significant spin-orbit splitting in the valence band.
The latter gives rise to two distinct excitonic features, labeled $A$ and $B$ in the schematic absorption spectrum shown in panel (c).
The primary excitonic features show a substantial binding energy, relative to the electron-hole continuum, e.g. for the neutral exciton, $E_A$.  
The exciton features exhibit a fine structure, with a splitting
attributable to the formation of trions labeled $A^-$ and $B^-$, with binding energies $E_{A^-}$ and $E_{B^-}$.

In this Letter we present a microscopic theory of excitonic effects in monolayer TMDs 
that describe the main features shown in  Fig.~\ref{fig:scheme} remarkably well.
Our theory is based on two-body and three-body excitonic Hamiltonians in the effective
mass approximation with screened interactions appropriate for strictly 2D semiconductors.
The Hamiltonians are parametrized by \textit{ab initio} calculations.
Variational wavefunctions, inspired by previous treatments of excitons in semiconductor quantum wells~\cite{ste97,ste98,ess00},
are employed.
By treating neutral and charged excitons on an equal footing, we achieve an internal consistency that
yields accurate, nontrivial predictions for neutral excitons while also providing quantitative insight into the
more complex trion species as well as trion binding energies that agree well with those inferred from experiment.
Our theory yields exciton binding energies in good, overall agreement with 
recent \textit{ab initio} calculations based on the Bethe-Salpeter
equation for TMDs~\cite{ram12,kom12,fen12,shi13}.
Extension of those calculations to the three-body trion problem is expected to be quite challenging.

Within the effective mass approximation, $\mu^{-1} = m_e^{-1} + m_h^{-1}$,
our theory employs the neutral excitonic Hamiltonian
\begin{equation}\label{eq:exciton_ham}
H_X = -\frac{\nabla^2_{\vrho}}{2\mu} - V_{2D}(\rho)
\end{equation}
and trion Hamiltonian
\begin{equation}\label{eq:trion_ham}
\begin{split}
H_{X^-} &= -\frac{1}{2\mu}\left(\nabla_{\vrho_1}^2+\nabla_{\vrho_2}^2\right) -\frac{1}{2m_h} \nabla_{\vrho_1}\cdot\nabla_{\vrho_2} \\
    &\hspace{1em} - V_{2D}(\rho_1) - V_{2D}(\rho_2) + V_{2D}(|\vrho_1-\vrho_2|),
\end{split}
\end{equation}
the latter of which is a generalization of the familiar Hamiltonian for the negative hydrogen ion~\cite{bet77} or for trions in quasi-2D quantum
wells~\cite{ste97,ste98,ess00}. 
It implicitly assumes that the trion can be
treated as an isolated, three-body problem 
reached in the low doping limit~\footnote{This approximation of isolated trions may be justified in the low doping limit due to disorder
induced localization of carriers on lengthscales comparable to the exciton and trion radii.}.
This approximation precludes
the observation of Fermi edge effects arising from the dynamical response of the electron gas~\cite{haw91,hua00}, an effect which has been observed
in the absorption spectra of MoS$_2$~\cite{mak13_trion}.
We also neglect interband mixing, due to the large spin-orbit splitting in TMDs, and consider only
the $A$ exciton and and its associated trion feature (see Fig.~\ref{fig:scheme}(c)); the $B$ features could be treated analogously.  We also
neglect any intervalley ($K$--$K^\prime$) coupling under the assumption of a selective, circularly polarized excitation~\cite{xia12,zen12,mak12}. The use
of linear polarization can excite coherent superpositions of valley excitons, inducing a valley exchange interaction~\cite{jon13}, also not treated here.

In typical experiments, the monolayer TMD material is surrounded by an environment 
with dielectric constants $\varepsilon_1$ (above) and $\varepsilon_2$
(below), but the electron and hole are restricted to orbitals that are primarily made up of TM $d$-states at the center of the trilayer TMD unit.
When there is a large dielectric contrast, which is typical of monolayer TMDs in vacuum or on weak dielectrics,
the effective in-plane 2D interaction for charges separated by $\rho = (x^2+y^2)^{1/2}$ reduces to a form derived by Keldysh~\cite{kel79},
\begin{equation}\label{eq:int}
V_{2D}(\rho) = \frac{\pi e^2}{(\varepsilon_1+\varepsilon_2)\rho_0} \left[ H_0\left(\frac{\rho}{\rho_0}\right) - Y_0\left(\frac{\rho}{\rho_0}\right) \right],
\end{equation}
where $H_0$ and $Y_0$ are the Struve function and the Bessel function of the second kind.
This interaction behaves like a screened $1/\rho$ Coulomb potential at long range, but has a weaker logarithmic divergence at
short range, where the crossover is determined by the screening length $\rho_0$.  The above interaction follows 
for a geometry which assumes the monolayer material has a thickness $d$ and isotropic dielectric constant $\varepsilon$,
for which the screening length is given by $\rho_0= d\varepsilon/(\varepsilon_1+\varepsilon_2)$.
In the strictly 2D limit of a polarizable plane in vacuum ($\varepsilon_{1,2}=1$), Cudazzo~\textit{et al.}~have recently
rederived Eq.~(\ref{eq:int}), showing that the screening length is 
given by $\rho_0 = 2\pi \chi_{2D}$, where $\chi_{2D}$ is the
2D polarizability of the planar material~\cite{cud11}.  
For the case of surrounding vacuum,
we have numerically verified that the screening length often times can be
accurately calculated using either definition of $\rho_0$, \textit{vide infra}, assuming
that the relevant dielectric constant of the monolayer is the in-plane component of the dielectric tensor of the \textit{bulk} material.
Here, we focus on freestanding monolayer TMDs,
but in future work on monolayer TMDs in novel
environments, the more general treatment of screening will be needed.

The necessary parameters for the exciton and trion Hamiltonians can be calculated from first principles.
The effective masses can be extracted from the low energy band structure (see Fig.~\ref{fig:scheme}(b)), 
calculated in density functional theory (DFT) or the $GW$ approximation~\cite{hyb86}.  
To extract the 2D polarizability, and thus the screening length $\rho_0$, we modify the protocol
in Ref.~\onlinecite{cud11} slightly.  We employ the relation
\begin{equation}\label{eq:eps_perp}
\varepsilon^\perp(L_c) = 1 + \frac{4\pi\chi_{2D}}{L_c} + O(1/L_c^2)
\end{equation}
where $L_c$ is the interlayer separation for a supercell containing \textit{two} AB-aligned monolayers
of TMD separated by vacuum.
The in-plane dielectric constant $\varepsilon^{\perp}$ is the $(q_x,q_y) \rightarrow 0$ limit of the head of the inverse dielectric tensor,
calculated within
the random phase approximation (RPA)~\cite{hyb87a}.
Our protocol naturally interpolates between bulk TMDs ($L_c = c/2$ where $c$ is the lattice constant) and
monolayer TMDs ($L_c \rightarrow \infty$).
This procedure tests the extraction of the monolayer 2D polarizability from the bulk dielectric constant via
Eq.~(\ref{eq:eps_perp}) retaining only the term of order $1/L_c$.

We study four monolayer TMDs: MoS$_2$, MoSe$_2$, WS$_2$, and WSe$_2$.  
The effective mass of the
electron and hole were calculated based on the parametrized band structures of Xiao \textit{et al.}~\cite{xia12}.  The $k \cdot p$ Hamiltonian
adopted in that work includes terms up to first order in $k$, 
yielding identical electron and hole masses. 
Higher order terms in $k$
predict differing effective masses~\cite{rost13,*kor13}, as also found in \textit{ab initio} calculations~\cite{pee12,ram12,che12,shi13}.
For evaluation of the polarizability, DFT and subsequent RPA calculations were performed
with the \textsc{quantum espresso}~\cite{gia09} and \textsc{berkeleygw}~\cite{des12} packages, respectively.
For MoS$_2$, in addition to the RPA result obtained with DFT input, we have also calculated
the RPA dielectric constant with an approximate $GW$ input, obtained by applying an $L_c$-dependent rigid shift to the unoccupied DFT bands,
$\Delta E_c^{GW}(L_c) = \Delta E_c^{GW}(\infty) - \alpha/L_c$, with $\Delta E_c^{GW}(\infty) = 1.2$ eV and $\alpha = 6.15$ eV$\cdot$\AA,
based on the results of Ref.~\onlinecite{kom12}.
Further computational details appear in the Supplemental Material.

\begin{figure}[t]
\centering
\includegraphics[scale=1.0]{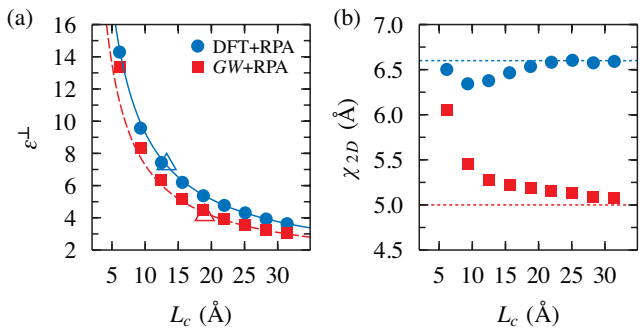}
\caption{In-plane dielectric constant (a) and two-dimensional polarizability (b) of MoS$_2$
as a function of the interlayer separation $L_c$ employed in the supercell calculations.  The smallest value
of $L_c$ employed corresponds to bulk MoS$_2$. Solid and dashed lines in panel (a) correspond to $1 + 4\pi \chi_{2D}/L_c$
with $\chi_{2D}$ extracted from panel (b).
Open symbols denote the values obtained via DFT and self-consistent $GW$ reported in Refs.~\onlinecite{mol11}
and \onlinecite{che12}, respectively.
}
\label{fig:eps_L}
\end{figure}

Using MoS$_2$ as an example, Fig.~\ref{fig:eps_L} shows the calculated 
dielectric constant $\varepsilon^{\perp}$ and the two-dimensional polarizability $\chi_{2D}$
as a function of the interlayer separation $L_c$ employed in the supercell calculations.  The dielectric constant clearly follows
Eq.~(\ref{eq:eps_perp}), giving the trivial limit of unity in the $L_c\rightarrow \infty$ limit. 
Many studies utilize similar 3D supercells to calculate dielectric properties
for 2D monolayer materials~\cite{sch11,mol11,che12,ram12}.
Two reported values for MoS$_2$~\cite{mol11,che12} are plotted in Fig.~\ref{fig:eps_L}(a),
showing agreement with the present results.  
Clearly, the dielectric constant at a fixed supercell size
together with an effective $1/\varepsilon r$ screened Coulomb interaction does not
represent dielectric screening in monolayer TMDs.
Furthermore, use of the conventional, 3D
Wannier-Mott theory with such a model to estimate exciton binding energies or radii~\cite{che12,ram12}
is not physical. 
In contrast, the two-dimensional polarizability
shown in Fig.~\ref{fig:eps_L}(b) converges to a finite and physically meaningful value 
independent of the final supercell size~\footnote{The convergence of the $GW$ polarizability
is much slower due to the changing
bandgap and required a $1/L_c \rightarrow 0$ extrapolation to the monolayer limit.}.
Specifically, we find $\chi_{2D} = 6.6$ \AA\ and 5.0 \AA, for DFT and $GW$, respectively. 
These values imply a two-dimensional screening length
of $\rho_0 \approx$ 30--40 \AA. 
To elucidate trends across materials,  
we use the DFT+RPA value and discuss the impact of the smaller $GW$ polarizability below.
Interestingly, we see that the DFT polarizability extracted from bulk MoS$2$ is extremely close to its
converged monolayer value, showing the near-equivalence of the two previously discussed definitions
of screening length,
$\rho_0 = 2\pi \chi_{2D}(L_c = c/2) = d (\varepsilon^\perp-1) /2 \approx d \varepsilon^\perp / 2$.

As a first approximation to the neutral excitonic properties of monolayer TMDs, we employ the total exciton wavefunction
$\Psi_X(\vr_e,\vr_h) = \psi_X(\vr_e-\vr_h) \phi_c(\vr_e) \phi_v(\vr_h)$, where $\phi_c$ and $\phi_v$ are conduction and valence Bloch wavefunctions,
with a simple
variational guess for the envelope function,
\begin{equation}\label{eq:exciton}
\psi_X(\vrho;a) \equiv \psi_X(\rho;a) = \sqrt{2/\pi a^2} \exp(-\rho/a).
\end{equation}
This variational wavefunction becomes the \textit{exact} ground state wavefunction in
the limit of weak screening, where $V(\rho) \rightarrow 1/\rho$. For a nonzero
polarizability, the wavefunction is no longer exact, but will exhibit the correct asymptotic behavior, i.e. exponential decay
for distances larger than the screening length $\rho_0$.
For this wavefunction, the kinetic energy is easily shown to be
$T(a) = 1/(2\mu a^2)$
and the potential energy $V(a)$ is readily evaluated by quadrature. 
The exciton binding energy is then found by minimizing $E_X = T(a)+V(a)$, where the optimum value of $a$ is an estimate of the exciton radius.

For the trion envelope
wavefunction, we consider the simple variational form 
\begin{equation}\label{eq:trion}
\begin{split}
\psi_{X^-}(\vrho_1,\vrho_2;a,b) &= 2^{-1/2} \big[ \psi_X(\rho_1;a)\psi_X(\rho_2;b) \\
    &\hspace{3em} + \psi_X(\rho_1;b)\psi_X(\rho_2;a) \big],
\end{split}
\end{equation}
a symmetrized product of exciton wavefunctions. 
First proposed by Chandrasekhar~\cite{cha44}, it is perhaps the only two-parameter
wavefunction to correctly predict a bound state of the negative hydrogen ion~\cite{bet77}.  The differing exciton radii, $a\neq b$, essentially
allows one electron to sit close to the hole, near the neutral exciton radius, while the other is further away to minimize the
unfavorable electron-electron repulsion.  A polarization term $(1+c\rho_{12})$ can also be included, although we will not do so here for simplicity.
For such a variational wavefunction, Eq.~(\ref{eq:trion}), with no dependence on the distance between the two
electrons, the so-called Hughes-Eckart term $\nabla_{\vrho_1}\cdot\nabla_{\vrho_2}$ vanishes~\cite{hog10}, simplifying the
numerical calculations.  Again, the kinetic energy can be evaluated analytically and the potential energy can be calculated numerically
as a three-dimensional integral.

The results for all four TMDs considered in this work 
are summarized in Table~\ref{tab:binding}.  Exciton binding energies are all predicted to be around 0.5 eV, with the ordering
MoS$_2$ $\gtrsim$ WS$_2$ $>$ MoSe$_2$ $\gtrsim$ WSe$_2$.
This trend generally agrees with recent \textit{ab initio}
Bethe-Salpeter equation (BSE) calculations on the same four materials~\cite{ram12}.  
Specifically for MoS$_2$, we find a binding energy of 0.54 eV and an exciton radius of 10.4 \AA.
Four recent BSE studies~\cite{ram12,kom12,fen12,shi13}, which vary in details of implementation, 
give results that vary by a factor of two, falling between 0.5 and 1.1 eV (Table~\ref{tab:binding}). 
Two technical challenges need to be fully resolved: convergence with respect to Brillouin zone sampling and the extrapolation
of the results to $L_c\rightarrow \infty$ limit, a particular challenge for the $GW$ results~\cite{kom12}.
Self-consistency would reduce screening, as is evident in Fig.~\ref{fig:eps_L}.
If we use our $GW$ polarizability in the monolayer limit, we find a
correspondingly larger binding energy of about 0.7 eV.  It is common for DFT to overestimate polarizability, and so taken together, our variational
estimates predict an exciton binding energy between 0.5 and 0.7 eV.  
All things considered, our variational estimate for the exciton binding energy is in good agreement with available \textit{ab initio} calculations.

\begin{table}[t]
\caption{Reduced mass (in $m_0$), polarizability (in \AA), exciton binding energies (in eV) and trion binding energies (in meV)
of TMDs as calculated with DFT+RPA. Many-body Bethe-Salpeter equation (BSE) exciton binding energies and experimental negative
trion binding energies are also listed.}
\begin{tabular*}{0.48\textwidth}{@{\extracolsep{\fill}} lcccccc }
\hline\hline
         &       &             & \multicolumn{2}{c}{Exciton binding energy}  & \multicolumn{2}{c}{Trion binding energy}     \\
\cline{4-5}\cline{6-7}
         & $\mu$ & $\chi_{2D}$ & Theory  & BSE                & Theory      & Exp  \\
\hline
MoS$_2$  & 0.25  & 6.60        & 0.54    & 1.03, 1.1~\cite{ram12,kom12} & 26          & 18 ~\cite{mak13_trion} \\
     &     &           &       &  0.5, 0.54~\cite{fen12,shi13} &             &    \\
MoSe$_2$ & 0.27  & 8.23        & 0.47    & 0.91~\cite{ram12}         & 21          & 30 ~\cite{ross13} \\
WS$_2$   & 0.16  & 6.03        & 0.50    & 1.04, 0.54 ~\cite{ram12,shi13} & 26          & \\
WSe$_2$  & 0.17  & 7.18        & 0.45    & 0.90~\cite{ram12}          & 22          & 30 ~\cite{jon13} \\
\hline\hline
\end{tabular*}
\label{tab:binding}
\end{table}

Carrying out the variational minimization of $E_{X^-} = \langle \psi_{X^-} | H_{X^-} | \psi_{X^-} \rangle$ for MoS$_2$, we find a trion binding energy (defined as
the difference between the trion and exciton variational energies) between 26 and 32 meV using the DFT and $GW$ polarizability,
respectively.  These values are impressively close to the experimental
value of 18 meV~\cite{mak13_trion}, suggesting that the approximations used here, including the form of the variational
wavefunction, are accurate and physically meaningful.
We find optimal radii of $a = 10.3$ \AA\ and
$b = 25.2$ \AA, i.e. one electron is at the neutral exciton radius while the other is more than twice as far away,
just as in the negative hydrogen ion.  The largeness of this trion binding energy, which is almost exactly equal to thermal energy
at room temperature, suggests that trions are intrinsically abundant and may play active roles in the excitonic physics of monolayer TMDs.

\begin{figure}[t]
\centering
\includegraphics[scale=0.4]{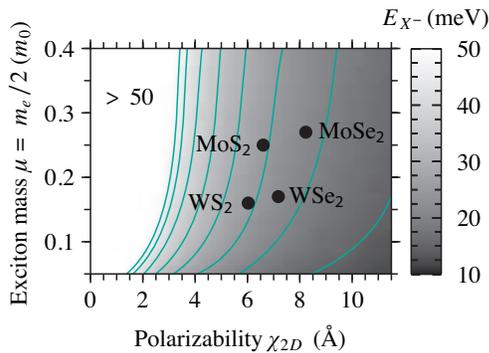}
\caption{
Calculated trion binding energy as a function of the exciton mass $\mu$ and the 2D polarizability $\chi_{2D}$, along with the
four TMDs considered in this work.  Contours are plotted in 5 meV increments.
It is evident why the sulfides and selenides each have essentially the same trion binding energies
despite differing material properties.
}
\label{fig:trion}
\end{figure}

The calculated trion binding energies for all four TMDs studied fall in the range of 20--30 meV, in reasonable agreement
with recently measured trion binding energies~\cite{mak13_trion,ross13,jon13}.  The similarity of trion binding energies in MoSe$_2$ and
WSe$_2$ is perfectly reproduced.
We find competing effects in the trion binding energy, parallel to the well-known trends for the exciton binding energy.
As Fig.~\ref{fig:trion} shows, increase in effective mass or reduction in polarizability both lead to stronger trion binding.
The exciton mass is largely determined by
the metal (i.e. W $5d$ versus Mo $4d$ electrons) whereas the polarizability depends on both the metal and the chalcogen: selenides have larger polarizabilities
than sulfides, and within a given chalcogenide family, molybdenum yields larger polarizabilities than tungsten.  
This argument also predicts a larger trion binding energy in
MoS$_2$ than in MoSe$_2$, contrary to the limited experimental results to
date~\cite{mak13_trion,ross13}.  However,
while the experiments on MoSe$_2$ and WSe$_2$ were done almost identically, the experiments on MoS$_2$ required significant gating to
achieve charge neutrality and exhibited extensive broadening in the lineshapes, both argued to be artifacts of defects introduced by
mechanical exfoliation~\cite{mak13_trion}.  Definitive trends for intrinsic trion binding energies remain an ongoing challenge for both theory
and experiment~\footnote{We have also investigated the use of Eq.~(\ref{eq:int}) with $\varepsilon_1 = 1$ and $\varepsilon_2 = 3.9$ to
simulate the SiO$_2$ dielectric substrate in experiments. This approach yields a trion binding energy in much better agreement
with experiment for MoS$_2$ but binding energies which are significantly underestimated for MoSe$_2$ and WSe$_2$. However, due to
imperfect monolayer-substrate contact geometries, any intermediate vacuum spacing will reduce the effective dielectric constant of the environment.}.

We briefly consider the positive trion.  Its Hamiltonian is identical to Eq.~(\ref{eq:trion_ham}), except that the electron mass replaces the hole mass in
the Hughes-Eckart term.  Since this term vanishes for our choice of wavefunction, we predict the positive trion binding energy
to be identical to that of the negative trion.  More generally, any difference in the electron
and hole masses only affects the binding energy to the extent that the true wavefunction depends explicitly on the distance between the electrons (for $X^-$)
or between the holes (for $X^+$).
This simple result may explain the equivalent positive and negative
trion binding energies recently observed in monolayer MoSe$_2$~\cite{ross13}, although more recent results on WSe$_2$ exhibit asymmetric
trion binding energies~\cite{jon13}.

While our variational approach has proved very effective, particularly to elucidate trends in the trion binding energies,
several physical effects remain to be quantified.
A non-variational treatment will obviously increase the neutral exciton binding energy.
Although we have neglected
the repulsive electron-hole exchange interaction, which would decrease the binding energy, the relatively large exciton
radius suggests that this contribution will be small.  
The trion binding energy, being an
energy \textit{difference}, is presumably less sensitive to these effects, such that a favorable cancellation
of errors is likely responsible for the observed accuracy as compared to recent experiments. This latter effect
is apparent in comparing binding energies based on DFT and $GW$ polarizabilities: while the exciton binding energy increases
by 40\%, the trion binding energy only increases by 20\%.
Other atomic-scale factors include local fields in the screened interaction at shorter range, the role of
the perpendicular extent of the electron and hole wavefunctions, and
a more accurate treatment of the low-energy band structure that accounts for anisotropy in the effective mass and trigonal warping effects~\cite{rost13,*kor13}.

To summarize, we have presented a simple, physically appealing theoretical treatment of both neutral and charged excitons in monolayers
of TMDs, a family of prototypical two-dimensional semiconductors.  
Our results highlight the strong effective Coulomb interactions
in monolayer TMDs and related 2D semiconductors that result in a dominant role
for excitons in the low energy optical physics, including bound trions that may be further 
engineered to play a significant role at room temperature for device applications.

We thank Jens Kunstmann, Eran Rabani, Tony Heinz, and Louis Brus for invaluable discussions.
This work was supported in part by the Center for Re-Defining Photovoltaic Efficiency through Molecule Scale
Control, an Energy Frontier Research Center funded by the U.S. Department of Energy, Office of Science, Office of
Basic Energy Sciences under Award Number DE-SC0001085.
This work was carried out in part at the Center for Functional Nanomaterials, Brookhaven National Laboratory,
which is supported by the U.S. Department of Energy, Office of Basic Energy Sciences under Contract No.~DE-AC02-98CH10886 (M.S.H).
T.C.B. was supported in part by the Department of Energy, Office of Science 
under Contract No.~DE-AC05-06OR23100.

\bibliography{mos2}

\clearpage
\onecolumngrid
\section{Supplemental Material}
\twocolumngrid

\section{Computational details}

DFT calculations were performed with the \textsc{quantum espresso}~\cite{gia09} software package,
using a $12\times 12\times n_{kz}$ $k$-point grid with $n_{kz}$ between 3 (for bulk) and 1 (for monolayer), using the exchange-correlation
functional of Perdew, Burke, and Ernzerhof~\cite{per96}, norm-conserving pseudopotentials, and a plane-wave cutoff of 40 Ry ($\sim$ 550 eV). RPA calculations
were done with the \textsc{berkeleygw}~\cite{des12} package on the same $k$-point grid and included 50 unoccupied bands.
The size of the dielectric matrix is determined by $G^2 < E_{\rm cut}$ where the
cutoff energy is equal to the energy of the highest unoccupied band included.  The $\vq\rightarrow 0$ limit is taken numerically
with a slightly shifted $k$-point grid as described in Ref.~\onlinecite{des12}.

For all materials studied, we employed experimental lattice constants and metal-chalcogen separations as given
in Table~\ref{tab:geom}.

\begin{table}[b]
\begin{tabular*}{0.48\textwidth}{@{\extracolsep{\fill}} lccc }
\hline\hline
          & $a$ (\AA) & $c$ (\AA) & $d_{MX}$ (\AA) \\
\hline
MoS$_2$   & 3.16      & 12.30     & 1.59           \\
MoSe$_2$  & 3.30      & 12.94     & 1.67           \\
WS$_2$    & 3.16      & 12.35     & 1.59           \\
WSe$_2$   & 3.29      & 12.98     & 1.67           \\
\hline\hline
\end{tabular*}
\caption{Crystal structure lattice constants ($a$ and $c$) and metal-chalcogen vertical separation ($d_{MX}$)
for the monolayer and bulk TMDs employed in this work.}
\label{tab:geom}
\end{table}

\section{Macroscopic dielectric constants of bulk MoS$_2$}

Unlike in the case of monolayer systems, the static dielectric constant is well-defined for bulk TMDs.  Using
the procedure described in the text, we have calculated the transverse and longitudinal dielectric
constant of bulk MoS$_2$ as an example.  These values are reported in Table~\ref{tab:bulk_eps} and compared to other
recent values found in the literature.

\begin{table}[b]
\begin{tabular*}{0.48\textwidth}{@{\extracolsep{\fill}} lrr }
\hline\hline
Reference           & $\varepsilon^{\perp}$ & $\varepsilon^{||}$ \\
\hline
Present work (PBE)             & 14.29                 & 6.87               \\
Present work (approximate $GW$)            & 13.36                 & 6.60               \\
\onlinecite{mol11} (LDA)       & 15.40                 & 7.43               \\
\onlinecite{che12} (sc$GW$)    & 13.5                  & 8.5                \\
\onlinecite{ram12} ($G_0W_0$)  & $\sim$ 14.5           &                    \\
\hline\hline
\end{tabular*}
\caption{Static dielectric constants of bulk MoS$_2$ as determined by a variety of methods in the literature.}
\label{tab:bulk_eps}
\end{table}

\end{document}